\documentclass[11pt,a4paper]{article}
\usepackage{amsfonts}
\usepackage{amsmath,amssymb}
\usepackage{epsfig,graphicx}

\input{tcilatex}

\begin{document}

\begin{center}
{\Large Non-Abelian Gauge Fields as Pseudo-Goldstone Vector
Bosons}

\bigskip

\textbf{J.L.~Chkareuli\ and \ J.G. Jejelava}

\bigskip

\textit{E. Andronikashvili} \textit{Institute of Physics and }

\textit{I. Chavchavadze State University, 0177 Tbilisi, Georgia\ \vspace{0pt}%
\\[0pt]
}

\bigskip \bigskip\bigskip \bigskip\bigskip \bigskip\bigskip \bigskip

\textbf{Abstract}

\bigskip
\end{center}

We argue that non-Abelian gauge fields can be treated as the
pseudo-Goldstone vector bosons caused by spontaneous Lorentz invariance
violation (SLIV). To this end, the SLIV\ which evolves in a general
Yang-Mills type theory with the nonlinear vector field constraint $Tr(%
\boldsymbol{A}_{\mu }\boldsymbol{A}^{\mu })=\pm M^{2}$\ ($M$ is a proposed
SLIV scale) imposed is considered in detail. With an internal symmetry group
$G$ having $D$ generators not only the pure Lorentz symmetry $SO(1,3)$, but
the larger accidental symmetry $SO(D,3D)$ of the SLIV constraint in itself
appears to be spontaneously broken as well. As a result, while the pure
Lorentz violation still generates only one genuine Goldstone vector boson,
the accompanying pseudo-Goldstone vector bosons related to the $SO(D,3D)$
breaking also come into play in the final arrangement of the entire
Goldstone vector field multiplet. Remarkably, they remain strictly massless,
being protected by gauge invariance of the Yang-Mills theory involved. We
show that, although this theory contains a plethora of Lorentz and $CPT$
violating couplings, they do not lead to physical SLIV effects which turn
out to be strictly cancelled in all the lowest order processes considered.
However, the physical Lorentz violation could appear if \ the internal gauge
invariance were slightly broken at very small distances influenced by
gravity. For the SLIV scale comparable with the Planck one the Lorentz
violation could become directly observable at low energies.

\thispagestyle{empty}\newpage

\section{Introduction}

The old idea\cite{bjorken} that spontaneous Lorentz invariance violation
(SLIV) may lead to an alternative theory of quantum electrodynamics still
remains extremely attractive in numerous theoretical contexts\cite{book}
(for some later developments, see the papers\cite{cfn}). The SLIV could
generally cause the appearance of massless vector Nambu-Goldstone modes
which are identified with photons and other gauge fields underlying the
modern particle physics framework like as Standard Model and Grand Unified
Theory. At the same time, the Lorentz violation by itself has attracted a
considerable attention in recent years as an interesting phenomenological
possibility appearing in various quantum field and string theories[4-9].

Early models realizing the SLIV conjecture were based on the four fermion
(current-current) interaction, where the proposed gauge field may appear as
a fermion-antifermion pair composite state\cite{bjorken}, in a complete
analogy with a massless composite scalar field in the original
Nambu-Jona-Lazinio model\cite{NJL}. Unfortunately, owing to the lack of a
starting gauge invariance in such models and composite nature of Goldstone
modes appeared it is hard to explicitly demonstrate that these modes really
form together a massless vector boson being a gauge field candidate.
Actually, one must make a precise tuning of parameters, including a
cancellation between terms of different orders in the $1/N$ expansion (where
$N$ is the number of fermion species involved), in order to achieve the
massless photon case\cite{suz}. Rather, there are in general three separate
massless Goldstone modes, two of which may mimic the transverse photons
polarizations, while the third one must properly be suppressed.

In this connection, the more instructive laboratory for SLIV consideration
proves to be some simple class of the QED type models having from the outset
a gauge invariant form, whereas the Lorentz violation is realized through
the nonlinear dynamical constraint imposed on the starting vector field $%
A_{\mu }$
\begin{equation}
\text{\ }A_{\mu }^{2}=n_{\mu }^{2}M^{2}\text{\ }  \label{cons1}
\end{equation}%
where $n_{\mu }$ is an properly oriented unit Lorentz vector, while $M$ is a
proposed SLIV scale. This constraint means in essence that the vector field $%
A_{\mu }$ develops the vacuum expectation value $\left\langle A_{\mu
}(x)\right\rangle $ $=n_{\mu }M$ and Lorentz symmetry $SO(1,3)$ breaks down
to $SO(3)$ or $SO(1,2)$ depending on the time-like ($n_{\mu }^{2}=+1$) or
space-like ($n_{\mu }^{2}=-1$) SLIV. Such QED model was first studied by
Nambu a long time ago\cite{nambu}, but only for the time-like SLIV case and
in the tree approximation. For this purpose he applied the technique of
nonlinear symmetry realizations which appeared successful in the handling of
the spontaneous breakdown of chiral symmetry in the nonlinear $\sigma $ model%
\cite{GL} and beyond\footnote{%
Actually, the simplest\ possible way to obtain the above supplementary
condition\ (\ref{cons1}) could be an inclusion the \textquotedblleft
standard\textquotedblright\ quartic vector field potential $V(A)=-\frac{%
m_{A}^{2}}{2}A_{\mu }^{2}+\frac{\lambda _{A}}{4}(A_{\mu }^{2})^{2}$ into the
QED type Lagrangian, as can be motivated to some extent\cite{alan} from the
superstring theory. This potential inevitably causes the spontaneous
violation of Lorentz symmetry in a standard way, much as an internal
symmetry violation is caused in a linear $\sigma $ model for pions\cite{GL}.
As a result, one has a massive Higgs mode (with mass $\sqrt{2}m_{A}$)
together with a massless Goldstone mode associated with photon. Furthermore,
just as in the pion model one can go from the linear model for the SLIV to
the non-linear one taking a limit $\lambda _{A}\rightarrow \infty ,$ $%
m_{A}^{2}\rightarrow \infty $ (while keeping the ratio $m_{A}^{2}/\lambda
_{A}$ to be finite). This immediately leads to the constraint (\ref{cons1})
for vector potential $A_{\mu }$ with $n_{\mu }^{2}M^{2}=m_{A}^{2}/\lambda
_{A}$, as it appears from a validity of its equation of motion. Another
motivation for the nonlinear vector field constraint (\ref{cons1}) might be
an attempt to avoid the infinite self-energy of the electron in a classical
electrodynamics, as was originally indicated by Dirac\cite{dir} and extended
later to various vector field theory cases\cite{vent}.}.

In the present paper, we mainly address ourselves\ to the Yang-Mills gauge
fields as the possible vector Goldstone modes (Sec.3) once some basic
ingredients of the Goldstonic QED model\ are established in a general SLIV\
case (Sec.2). This problem has been discussed many times in the literature
within quite different contexts, such as the Yang-Mills gauge fields as the
Goldstone modes for the spontaneous breaking of general covariance in a
higher-dimensional space\cite{freund} or for the nonlinear realization of
some special infinite parameter gauge group\cite{ogi2}. However, all these
considerations look rather speculative and optional. Specifically, they do
not give a correlation between the SLIV induced photon case, from the one
hand, and the Yang-Mills gauge field\ case, from the other. In contrast, our
approach is solely based on the spontaneous Lorentz violation thus properly
generalizing the Nambu's QED model\cite{nambu} to the non-Abelian internal
symmetry case. Just in this approach evolved the interrelation between both
of cases appears most transparent. We will see that in the Yang-Mills theory
case with an internal symmetry group\ $G$ having $D$ generators not only the
pure Lorentz symmetry part $SO(1,3)$ in the symmetry $SO(1,3)\times G$ of
the Lagrangian, but the larger accidental symmetry $SO(D,3D)$ of the SLIV
constraint $Tr(\boldsymbol{A}_{\mu }\boldsymbol{A}^{\mu })=\pm M^{2}$\ in
itself is spontaneously broken as well. Because the starting non-Abelian
theory proves to be expanded about the vacuum which violates the much higher
accidental symmetry appeared, many extra massless modes, the
pseudo-Goldstone vector bosons (PGB), have to arise. Actually, while the
spontaneous Lorentz violation on its own still generates only one genuine
Goldstone vector boson, the accompanying vector PGBs related to the $%
SO(D,3D) $ breaking also come into play in the final arrangement of the
entire Goldstone vector field multiplet. Remarkably, in contrast to the
familiar scalar PGB case\cite{GL} the vector PGBs remain strictly massless
being protected by the non-Abelian gauge invariance of the Yang-Mills theory
involved. Then in Sec.4 we show by some examples of the lowest order SLIV
processes that, while the Goldstonic non-Abelian theory evolved contains a
rich variety of Lorentz and $CPT$ violating couplings, it proves to be
physically indistinguishable from a conventional Yang-Mills theory.
Actually, one of the goals of the present work is to explicitly demonstrate
that a conventional Yang-Mills theory (as well as QED) is in fact the
spontaneously broken theory. The Lorentz violation, due to the quadratic
field constraint of the type\ (\ref{cons1}), renders this theory highly
nonlinear in the Goldstone vector modes, while physically equivalent to the
usual one. So, as well as in the pure QED case, the SLIV only means the
noncovariant gauge choice in the otherwise gauge invariant and Lorentz
invariant Yang-Mills theory. However, even a tiny breaking of the starting
gauge invariance at very small distances influenced by gravity would render
the SLIV physically significant. For the SLIV scale comparable with the
Planck one the spontaneous Lorentz violation could become directly
observable at low energies. We summarize the results obtained in the final
Sec.5.

\section{Goldstonic quantum electrodynamics}

The simplest SLIV model is given by a conventional QED Lagrangian for the
charged fermion field $\psi $
\begin{equation}
L(A,\psi )=-\frac{1}{4}F_{\mu \nu }F^{\mu \nu }+\overline{\psi }(i\gamma
\cdot \partial -m)\psi -eA_{\mu }\overline{\psi }\gamma ^{\mu }\psi
\label{lagr1}
\end{equation}%
where the nonlinear vector field constraint (\ref{cons1}) is imposed\cite%
{nambu}. For the resulting Lorentz violation, one can rewrite the Lagrangian
$L(A,\psi )$ in terms of \ the standard parametrization for the vector
potential $A_{\mu }$
\begin{equation}
A_{\mu }=a_{\mu }+\frac{n_{\mu }}{n^{2}}(n\cdot A)\text{ \ \ \ \ \ \ }%
(n^{2}\equiv n_{\mu }^{2})  \label{par}
\end{equation}%
where the $a_{\mu }$ is pure Goldstonic mode
\begin{equation}
\text{\ }n\cdot a=0\text{\ }
\end{equation}%
while the effective Higgs mode (or the $A_{\mu }$ component in the vacuum
direction) is given according to the above nonlinear constraint (\ref{cons1}%
) by
\begin{equation}
\text{\ }n\cdot A\text{\ }=(M^{2}-n^{2}a_{\nu }^{2})^{\frac{1}{2}}=M-\frac{%
n^{2}a_{\nu }^{2}}{2M}+O(1/M^{2})  \label{constr1}
\end{equation}%
where, for definiteness, the positive sign for the above square root was
taken when expanding it in powers of $a_{\nu }^{2}/M^{2}$. Putting the
parametrization (\ref{par}) with the SLIV constraint (\ref{cons1}, \ref%
{constr1}) into our basic gauge invariant Lagrangian (\ref{lagr1}) one comes
to the truly Goldstonic model for QED. This model might look unacceptable
due to the inappropriately large Lorentz violating fermion bilinear $eM%
\overline{\psi }(\gamma \cdot n)\psi $ stemming from the vector-fermion
current interaction $eA_{\mu }\overline{\psi }\gamma ^{\mu }\psi $ in the
Lagrangian $L$ (\ref{lagr1}) when the expansion (\ref{constr1}) is taken.
However, thanks to a local invariance of the Lagrangian $L$ this term can be
gauged away by a suitable redefinition of the fermion field
\begin{equation}
\psi \rightarrow e^{ieM(n\cdot x)}\psi  \label{aw}
\end{equation}%
after which the above fermion bilinear is exactly cancelled by an analogous
term stemming from the fermion kinetic term. So, one eventually comes to the
essentially nonlinear SLIV Lagrangian for the Goldstonic $a_{\mu }$ field of
the type (taken in the first approximation in $a_{\nu }^{2}/M^{2}$)
\begin{eqnarray}
L(a,\psi ) &=&-\frac{1}{4}f_{\mu \nu }f^{\mu \nu }-\frac{1}{2}\delta (n\cdot
a)^{2}-\frac{1}{4}f_{\mu \nu }h^{\mu \nu }\frac{n^{2}a_{\rho }^{2}}{M}+
\label{NL} \\
&&+\overline{\psi }(i\gamma \cdot \partial +m)\psi -ea_{\mu }\overline{\psi }%
\gamma ^{\mu }\psi +\frac{en^{2}a_{\rho }^{2}}{2M}\overline{\psi }(\gamma
\cdot n)\psi  \notag
\end{eqnarray}%
We denoted its strength tensor by $f_{\mu \nu }=\partial _{\mu }a_{\nu
}-\partial _{\nu }a_{\mu }$, while $h^{\mu \nu }=n^{\mu }\partial ^{\nu
}-n^{\nu }\partial ^{\mu }$ is a new SLIV oriented differential tensor. This
tensor $h^{\mu \nu }$ acts on the infinite series in $a_{\rho }^{2}$ coming
from the expansion of the effective Higgs mode (\ref{constr1}) from which
the first order term $-n^{2}a_{\nu }^{2}/2M$ was only taken in this
expansion throughout the Lagrangian $L(a,\psi )$. Also, we explicitly
included the orthogonality condition $n\cdot a=0$ into Lagrangian through
the term which can be treated as the gauge fixing term (taking the limit $%
\delta \rightarrow \infty $) and retained the former notation for the
fermion $\psi $.

The Lagrangian (\ref{NL}) completes the Goldstonic QED construction for the
charged fermion field $\psi $. The model, as one can see, contains the
massless Goldstone modes given by the tree broken generators of the Lorentz
group, while keeping the massive Higgs mode frozen. These modes, lumped
together, constitute a single Goldstone vector boson associated with photon%
\footnote{%
Strictly speaking one can no longer use the standard definition of photon as
a state being the spin-1 representation of the (now spontaneously broken)
Poincare group. However, due to gauge symmetry of \ the starting QED
Lagrangian\ (\ref{lagr1}) the separate SLIV Goldstone modes appear combined
in such a way that a standard photon (taken in an axial gauge (4)) emerges.}%
. In the limit $M\rightarrow \infty $ the model is indistinguishable from a
conventional QED taken in the general axial (temporal or pure axial) gauge.
So, for this part of the Lagrangian $L(a,\psi )$ given by the zero-order
terms in $1/M$ the spontaneous Lorentz violation only means the noncovariant
gauge choice in otherwise the gauge invariant (and Lorentz invariant)
theory. Remarkably, furthermore, also all the other (first and higher order
in $1/M$) terms in the $L(a,\psi )$ \ (\ref{NL}), though being by themselves
the Lorentz and $CPT$ violating ones, do not lead to the physical SLIV
effects which turn out to be strictly cancelled in all the physical
processes involved. So, the nonlinear constraint (\ref{cons1}) imposed on
the standard QED Lagrangian (\ref{lagr1}) appears, in fact, as a possible
gauge choice, while the $S$-matrix remains unaltered under such a gauge
convention. This conclusion was first reached at tree level\cite{nambu} and
recently extended to the one-loop approximation\cite{ac}. All the one-loop
contributions to the photon-photon, photon-fermion and fermion-fermion
interactions violating the physical Lorentz invariance were shown to be
exactly cancelled as well. This means that the vector field constraint $%
A_{\mu }^{2}=n_{\mu }^{2}M^{2}$ which has been treated as the nonlinear
gauge choice at a tree (classical) level, remains just as a pure gauge
condition when quantum effects are also taken into account. Remarkably, this
conclusion appears to work also for a general Abelian theory case\cite{ck},
particularly, when the internal $U(1)$ charge symmetry is spontaneously
broken hand in hand with the Lorentz one. As a result, the massless photon
being first generated by the Lorentz violation become then massive due to
the standard Higgs mechanism, while the SLIV condition in itself remains to
be a gauge choice\footnote{%
Note in this connection that there was discussed\cite{nambu} a possibility
of an explicit construction of the gauge function corresponding to the
nonlinear gauge constraint (1) that would eliminate the need for all the
kinds of checks of gauge invariance mentioned above. Remarkably, the
equation for this gauge function appears to be mathematically equivalent to
the classical Hamilton-Jacobi equation of motion for a charged particle.
Thus, this gauge function should in principle exist because there is a
solution to the classical problem. However, this formal analogy only works
for the time-like SLIV ($n_{\mu }^{2}=+1$) in the pure QED leaving aside a
general Abelian theory when the gauge invariance can spontaneously be
broken. Apart from that, it does not generally extend to the non-Abelian
case (see next Section).
\par
{}}.

\section{Goldstonic Yang-Mills theory}

In this section, we extend our discussion to the non-Abelian internal
symmetry case given by a general group $G$ with generators $t^{i}$($%
[t^{i},t^{j}]=ic^{ijk}t^{k}$ and $Tr(t^{i}t^{j})=\delta ^{ij}$ where $%
c^{ijk} $ are structure constants and $i,j,k=0,1,...,D-1$). The
corresponding vector fields\ which transform according to its adjoint
representation are given in the proper matrix form $\boldsymbol{A}_{\mu
}=A_{\mu }^{i}t^{i}$, while the matter fields (fermions, for definiteness)
are presented in the fundamental representation column $\mathfrak{\psi }^{r}$
($r=0,1,...,d-1$) of $G$. By analogy with the above Goldstonic QED case we
take for them a conventional Yang-Mills type Lagrangian
\begin{equation}
\mathfrak{L}(\boldsymbol{A},\mathfrak{\psi })=-\frac{1}{4}\,Tr(\boldsymbol{F}%
_{\mu \nu }\boldsymbol{F}^{\mu \nu })+\overline{\mathfrak{\psi }}(i\gamma
\cdot \partial -m)\mathfrak{\psi }+g\overline{\mathfrak{\psi }}\boldsymbol{A}%
_{\mu }\gamma ^{\mu }\mathfrak{\psi }  \label{nab}
\end{equation}%
(where $\boldsymbol{F}_{\mu \nu }\boldsymbol{~=~}\partial _{\mu }\boldsymbol{%
A}_{\nu }-\partial _{\nu }\boldsymbol{A}_{\mu }-ig[\boldsymbol{A}_{\mu },%
\boldsymbol{A}_{\nu }]$ and $g$ stands for the universal coupling constant
in the theory) with the nonlinear SLIV constraint
\begin{equation}
Tr(\boldsymbol{A}_{\mu }\boldsymbol{A}^{\mu })=\mathfrak{n}_{\mu }^{2}M^{2},%
\text{ \ \ }\mathfrak{n}_{\mu }^{2}=\pm 1  \label{CON}
\end{equation}%
imposed\footnote{%
As in the Abelian case, the existence of such a constraint could be related
with some nonlinear $\sigma $ type SLIV\ model proposed for the vector field
multiplet $A_{\mu }^{i}$ in the Yang-Mills theory (\ref{nab}). Note in this
connection that, due to its generic antisymmetry, the familiar quadrilinear
terms $-\frac{1}{4}g^{2}Tr([\boldsymbol{A}_{\mu }\boldsymbol{,A}_{\nu
}])^{2} $ in the Lagrangian (\ref{nab}) do not contribute into the SLIV
since they identically vanish for any single-valued vacuum configuration $%
\left\langle A_{\mu }^{i}\right\rangle $.}. \ One can easily see that,
although we propose only the $SO(1,3)\times G$ invariance in the theory, the
SLIV constraint taken (\ref{CON}) possesses, in fact, the much higher
accidental symmetry $SO(D,3D)$ determined by the dimensionality $D$ of the $%
G $ group adjoint representation to which the vector fields $A_{\mu }^{i}$
are belonged. This symmetry is indeed spontaneously broken at a scale $M$
\begin{equation}
\left\langle A_{\mu }^{i}(x)\right\rangle \text{ }=\mathfrak{n}_{\mu }^{i}M
\label{vev}
\end{equation}%
with the vacuum direction given now by the `unit' rectangular matrix $%
\mathfrak{n}_{\mu }^{i}$ which describes both of the generalized SLIV cases
at once, time-like ($SO(D,3D)$ $\rightarrow SO(D-1,3D)$) or space-like ($%
SO(D,3D)$ $\rightarrow SO(D,3D-1)$), respectively, depending on the sign of
the $\mathfrak{n}_{\mu }^{2}\equiv \mathfrak{n}_{\mu }^{i}\mathfrak{n}^{\mu
,i}=\pm 1$. This matrix has only one non-zero element for both of cases
determined by the proper $SO(D,3D)$ rotation. They are, particularly, $%
\mathfrak{n}_{0}^{0}$ or $\mathfrak{n}_{3}^{0}$ provided that the vacuum
expectation value (\ref{vev}) is developed along the $i=0$ direction in the
internal space and along the $\mu =0$ or $\mu =3$ direction, respectively,
in the Minkowskian space-time. In response to each of these two breakings,
side by side with one true vector Goldstone boson and the $D-1$ scalar
Goldstone bosons corresponding to the spontaneous violation of actual $%
SO(1,3)\otimes G$ symmetry of the total Lagrangian $\mathfrak{L}$, \ the $%
D-1 $ vector pseudo-Goldstone bosons related to breaking of the accidental $%
SO(D,3D)$ symmetry of the SLIV constraint taken (\ref{CON}) are also
produced. Remarkably, in contrast to the familiar scalar PGB case\cite{GL}
the vector PGBs remain strictly massless being protected by the non-Abelian
gauge invariance of the starting Lagrangian (\ref{nab}). Together with the
aforementioned true vector Goldstone boson they complete the entire
Goldstonic vector field multiplet of the internal symmetry group $G$.

As in the Abelian case, upon an explicit use of the corresponding SLIV
constraint (\ref{CON}) being so far the only supplementary condition for
vector field multiplet $A_{\mu }^{i}$, one comes to the pure Goldstone field
modes $a_{\mu }^{i}$ identified in a similar way
\begin{equation}
\text{\ \ }A_{\mu }^{i}=a_{\mu }^{i}+\frac{\mathfrak{n}_{\mu }^{i}}{%
\mathfrak{n}^{2}}(\mathfrak{n}\cdot A)\text{ },\text{ \ }\mathfrak{n}\cdot
a\equiv \mathfrak{n}_{\mu }^{i}a^{\mu ,i}\text{\ }=0\text{ \ \ \ \ \ \ }(%
\mathfrak{n}^{2}\equiv \mathfrak{n}_{\mu }^{2})\text{ },  \label{sup'}
\end{equation}%
At the same time, an effective Higgs mode (i.e., the $A_{\mu }^{i}$
component in the vacuum direction $\mathfrak{n}_{\mu }^{i}$) is given by the
product $\mathfrak{n}\cdot A\equiv \mathfrak{n}_{\mu }^{i}A^{\mu ,i}$
determined by the SLIV constraint
\begin{equation}
\text{\ }\mathfrak{n}\cdot A\text{\ }=\left[ M^{2}-\mathfrak{n}^{2}(a_{\nu
}^{i})^{2}\right] ^{\frac{1}{2}}=M-\frac{\mathfrak{n}^{2}(a_{\nu }^{i})^{2}}{%
2M}+O(1/M^{2})  \label{constr''}
\end{equation}%
where, as earlier in the Abelian case, we took the positive sign for the
square root when expanding it in powers of $(a_{\nu }^{i})^{2}/M^{2}$. Note
that the general Goldstonic modes $a_{\mu }^{i}$, apart from pure vector
fields, contain the $D-1$ scalar ones, $a_{0}^{i^{\prime }}$ and $%
a_{3}^{i^{\prime }}$ ($i^{\prime }=1...D-1$), for the time-like ($\mathfrak{n%
}_{\mu }^{i}=\mathfrak{n}_{0}^{0}g_{\mu 0}\delta ^{i0}$) and space-like ($%
\mathfrak{n}_{\mu }^{i}=\mathfrak{n}_{3}^{0}g_{\mu 3}\delta ^{i0}$) SLIV,
respectively. They can be eliminated from the theory if one puts the proper
supplementary conditions on the $a_{\mu }^{i}$ fields which were still the
constraint free. Using their overall orthogonality (\ref{sup'}) to the
physical vacuum direction $\mathfrak{n}_{\mu }^{i}$ one can formulate these
supplementary conditions in terms of a general axial gauge for the entire $%
a_{\mu }^{i}$ multiplet%
\begin{equation}
n\cdot a^{i}\equiv n_{\mu }a^{\mu ,i}=0,\text{ \ }i=0...D-1  \label{sup''}
\end{equation}%
where $n_{\mu }$ is the unit Lorentz vector introduced in the Abelian case
which is now oriented in Minkowskian space-time so as to be parallel to the
vacuum matrix $\mathfrak{n}_{\mu }^{i}$. For such a choice the simple
equation holds%
\begin{equation}
\mathfrak{n}_{\mu }^{i}=s^{i}n_{\mu }\text{ \ \ \ }(s^{i}\equiv \frac{n\cdot
\mathfrak{n}^{i}}{n^{2}})  \label{id}
\end{equation}%
which shows that the rectangular vacuum matrix $\mathfrak{n}_{\mu }^{i}$ has
the factorized "two-vector" form. As a result, apart from the Higgs mode
excluded earlier by the orthogonality condition (\ref{sup'}), all the scalar
fields also appear eliminated, and only pure vector fields, $a_{\mu ^{\prime
}}^{i}$ ($\mu ^{\prime }=1,2,3$) or $a_{\mu ^{\prime \prime }}^{i}$ ($\mu
^{\prime \prime }=0,1,2$) for time-like or space-like SLIV, respectively,
are only left in the theory.

We now show that the such constrained Goldstone vector fields $a_{\mu }^{i}$
(with the supplementary conditions (\ref{sup''}) taken) appear truly
massless when the starting non-Abelian Lagrangian $\mathfrak{L}$ (\ref{nab})
is rewritten in the form determined by the SLIV. Actually, putting the
parametrization (\ref{sup'}) with the SLIV constraint (\ref{constr''}) into
the Lagrangian (\ref{nab}) one is led to the highly nonlinear Yang-Mills
theory in terms of the pure Goldstonic gauge field modes $a_{\mu }^{i}$.
However, as in the above Abelian case, one should first gauge away (using
the local invariance of the Lagrangian $\mathfrak{L}$) the enormously large,
while false, Lorentz violating terms appearing in the theory in the form of
the fermion and vector field bilinears. As one can readily see, they stem
from the couplings $g\overline{\mathfrak{\psi }}\boldsymbol{A}_{\mu }\gamma
^{\mu }\mathfrak{\psi }$ \ and $-\frac{1}{4}g^{2}Tr([\boldsymbol{A}_{\mu }%
\boldsymbol{,A}_{\nu }])^{2}$, respectively, when the effective Higgs mode
expansion (\ref{constr''}) is taken in the Lagrangian (\ref{nab}). Making
the appropriate redefinitions of the fermion ($\mathfrak{\psi }$ ) and
vector ($\boldsymbol{a}_{\mu }\equiv a_{\mu }^{i}t^{i}$) field multiplets
\begin{equation}
\mathfrak{\psi }\rightarrow U(\omega )\mathfrak{\psi }\text{ },\text{ \ \ }%
\boldsymbol{a}_{\mu }\rightarrow U(\omega )\boldsymbol{a}_{\mu }U(\omega )^{%
{\Huge \dagger }},\text{ \ }U(\omega )=e^{igM(\mathfrak{n}^{i}\cdot x)t^{i}}
\label{red1}
\end{equation}%
and using the evident equalities for the linear (in coordinate)
transformations $U(\omega )$ with the single-valued vacuum matrix $\mathfrak{%
n}_{\mu }^{i}$ ($\mathfrak{n}_{0}^{0}$ or $\mathfrak{n}_{3}^{0}$ for the
particular SLIV cases)
\begin{equation}
\partial _{\mu }U(\omega )=ig\mathfrak{n}_{\mu }^{i}t^{i}U(\omega
)=igU(\omega )\mathfrak{n}_{\mu }^{i}t^{i}
\end{equation}%
one can confirm that the abovementioned Lorentz violating terms are exactly
cancelled with the analogous bilinears stemming from their kinetic terms.
So, the final Lagrangian for the Goldstonic Yang-Mills theory takes the form
(in the first approximation in $(a_{\nu }^{i})^{2}/M^{2}$)
\begin{eqnarray}
\mathfrak{L}(a\boldsymbol{,}\mathfrak{\psi }) &=&-\frac{1}{4}Tr(\boldsymbol{f%
}_{\mu \nu }\boldsymbol{f}^{\mu \nu })-\frac{1}{2}\delta (n\cdot a^{i})^{2}+%
\frac{1}{4}Tr(\boldsymbol{f}_{\mu \nu }\boldsymbol{h}^{\mu \nu })\frac{%
\mathfrak{n}^{2}(a_{\nu }^{i})^{2}}{M}+  \notag \\
&&+\overline{\mathfrak{\psi }}(i\gamma \cdot \partial -m)\mathfrak{\psi }+g%
\overline{\mathfrak{\psi }}\boldsymbol{a}_{\mu }\gamma ^{\mu }\mathfrak{\psi
}-\frac{g\mathfrak{n}^{2}(a_{\nu }^{i})^{2}}{2M}\overline{\mathfrak{\psi }}%
(\gamma \cdot \text{\ }\mathfrak{n}^{k})t^{k}\mathfrak{\psi }  \label{nab3}
\end{eqnarray}%
where the tensor $\boldsymbol{f}_{\mu \nu }$ is, as usual, $\boldsymbol{f}%
_{\mu \nu }\boldsymbol{~=~}\partial _{\mu }\boldsymbol{a}_{\nu }-\partial
_{\nu }\boldsymbol{a}_{\mu }-ig[\boldsymbol{a}_{\mu },\boldsymbol{a}_{\nu }]$%
, while $\boldsymbol{h}_{\mu \nu }$ is a new SLIV oriented tensor of the
type
\begin{equation}
\boldsymbol{h}_{\mu \nu }=\boldsymbol{n}_{\mu }\partial _{\nu }-\boldsymbol{n%
}_{\nu }\partial _{\mu }+ig([\boldsymbol{n}_{\mu },\boldsymbol{a}_{\nu }]-[%
\boldsymbol{n}_{\nu },\boldsymbol{a}_{\mu }]),\text{ \ \ }\boldsymbol{n}%
_{\mu }\equiv \mathfrak{n}_{\mu }^{k}t^{k}
\end{equation}%
This tensor $\boldsymbol{h}_{\mu \nu }$ acts on the infinite series in $%
(a_{\nu }^{i})^{2}$ coming from the expansion of the effective Higgs mode (%
\ref{constr''}) from which only the first order term $-\mathfrak{n}%
^{2}(a_{\nu }^{i})^{2}/2M$ was taken throughout the Lagrangian $\mathfrak{L}%
(a\boldsymbol{,}\mathfrak{\psi })$. We also retained the former notations
for the fermion and vector field multiplets after transformations (\ref{red1}%
), \ and explicitly included the (axial) gauge fixing term into Lagrangian
according to the supplementary conditions taken (\ref{sup''}).

The theory derived gives a proper generalization of the nonlinear QED model%
\cite{nambu} for the non-Abelian case. It contains the massless vector boson
multiplet $a_{\mu }^{i}$ (consisting of one Goldstone and $D-1$
pseudo-Goldstone vector states) which gauges the starting internal symmetry $%
G$. In the limit $M\rightarrow \infty $ \ it is indistinguishable from a
conventional Yang-Mills theory taken in the general axial gauge. So, for
this part of the Lagrangian $\mathfrak{L}(a\boldsymbol{,}\mathfrak{\psi })$
given by the zero-order in $1/M$ terms the spontaneous Lorentz violation
only means the noncovariant gauge choice in the otherwise gauge invariant
(and Lorentz invariant) theory. However, one may expect that, just as it
appears in the nonlinear QED model, also all the first and higher order in $%
1/M$ terms in the $\mathfrak{L}$\ (\ref{nab3}), though being by themselves
the Lorentz and $CPT$ violating ones, do not lead to the physical SLIV
effects due to the mutual cancellation of their contributions into all the
physical processes appeared.

\section{The lowest order SLIV processes}

Let us now show that the simple tree level calculations related to the
Lagrangian $\mathfrak{L}(a\boldsymbol{,}\mathfrak{\psi })$ confirms in
essence this proposition. As an illustration, we consider SLIV processes in
the lowest order in $g$ and $1/M$ \ being the fundamental parameters of the
Lagrangian (\ref{nab3}). They are, as one can readily see, the
vector-fermion and vector-vector elastic scattering going in the order $g/M$%
, which we turn to once the Feynman rules in the Goldstonic Yang-Mills
theory are established.

\subsection{Feynman rules}

The corresponding Feynman rules, apart from the ordinary Yang-Mills theory
rules for

(i) the vector-fermion vertex $\ $%
\begin{equation}
-ig~\gamma _{\mu }~t^{i}  \label{ver3}
\end{equation}

(ii) the vector field propagator (taken in a general axial gauge $n^{\mu
}a_{\mu }^{i}=0$)
\begin{equation}
D_{\mu \nu \hspace*{0.05in}\hspace*{0.05in}\hspace*{0.05in}}^{ij}\left(
k\right) =-\frac{i\delta ^{ij}}{k^{2}}~\left( g_{\mu \nu }-\frac{n_{\mu
}k_{\nu }+k_{\mu }n_{\nu }}{n\cdot k}+\frac{n^{2}k_{\mu }k_{\nu }}{(n\cdot
k)^{2}}\right)  \label{prop1}
\end{equation}%
which automatically satisfies the orthogonality condition $n^{\mu }D_{\mu
\nu }^{ij}(k)=0$ and on-shell transversality $k_{\mu }D_{\mu \nu }^{ij}(k)=0$
$\ $ ($k^{2}=0$); the latter means that free vector fields with polarization
vector $\epsilon _{\mu }^{i}(k,k^{2}=0)$ are always appeared transverse $%
k^{\mu }\epsilon _{\mu }^{i}(k)=0$;

(iii) the 3-vector vertex (with vector field 4-momenta $k_{1},$ $k_{2}$ and $%
k_{3}$; all 4-momenta in vertexes are taken ingoing throughout) $\ \ $%
\begin{equation}
gc^{ijk}[(k_{1}-k_{2})_{\gamma }g_{\alpha \beta }+(k_{2}-k_{3})_{\alpha
}g_{\beta \gamma }+(k_{3}-k_{1})_{\beta }g_{\alpha \gamma }]  \label{ver0}
\end{equation}%
\ include the new ones, violating Lorentz and $CPT$ invariance, for \ \ \ \

(iv) the contact 2-vector-fermion vertex $\ \ $%
\begin{equation}
i\frac{g\mathfrak{n}^{2}}{M}(\gamma \cdot \mathfrak{n}^{k})\tau ^{k}g_{\mu
\nu }~\delta ^{ij}  \label{ver1}
\end{equation}

(v) another 3-vector vertex
\begin{equation}
-\frac{i\mathfrak{n}^{2}}{M}\left[ (k_{1}\cdot \mathfrak{n}^{i})k_{1,\alpha
}g_{\beta \gamma }\delta ^{jk}+(k_{2}\cdot \mathfrak{n}^{j})k_{2,\beta
}g_{\alpha \gamma }\delta ^{ki}+(k_{3}\cdot \mathfrak{n}^{k})k_{3,\gamma
}g_{\alpha \beta }\delta ^{ij}\right]  \label{ver2}
\end{equation}%
where the second index in the vector field 4-momenta $k_{1},$ $k_{2}$ and $%
k_{3}$ denotes their Lorentz components; \ \ \ \

(vi) the extra 4-vector vertex (with the vector field 4-momenta $k_{1,2,3,4}$
and their proper differences $k_{12}\equiv k_{1}-k_{2}$ \ etc.)%
\begin{eqnarray}
&&-\frac{\mathfrak{n}^{2}g}{M}[c^{ijp}\delta ^{kl}g_{\alpha \beta }g_{\gamma
\delta }(\mathfrak{n}^{p}\cdot k_{12})+c^{klp}\delta ^{ij}g_{\alpha \beta
}g_{\gamma \delta }(\mathfrak{n}^{p}\cdot k_{34})+  \notag \\
&&+c^{ikp}\delta ^{jl}g_{\alpha \gamma }g_{\beta \delta }(\mathfrak{n}%
^{p}\cdot k_{13})+c^{jlp}\delta ^{ik}g_{\alpha \gamma }g_{\beta \delta }(%
\mathfrak{n}^{p}\cdot k_{24})+  \label{ver4} \\
&&+c^{ilp}\delta ^{jk}g_{\alpha \delta }g_{\beta \gamma }(\mathfrak{n}%
^{p}\cdot k_{14})+c^{jkp}\delta ^{il}g_{\alpha \delta }g_{\beta \gamma }(%
\mathfrak{n}^{p}\cdot k_{23})]  \notag
\end{eqnarray}%
where only the terms which can not lead to contractions of the rectangular
vacuum matrix $\mathfrak{n}_{\mu }^{p}$ with vector field polarization
vectors $\epsilon _{\mu }^{i}(k)$ are presented. These contractions are in
fact vanished due to the gauge taken (\ref{sup''}), $\mathfrak{n}^{p}\cdot
\epsilon ^{i}=$ $s^{p}(n\cdot \epsilon ^{i})=0$ (with a factorized
two-vector form for the matrix $\mathfrak{n}_{\mu }^{p}$ (\ref{id}) used).

Just the rules (i-vi) are needed to calculate the lowest order amplitudes of
the processes we have mentioned in the above.

\subsection{Vector boson scattering on fermion}

This process is directly related to two SLIV diagrams one of which is given
by the contact $a^{2}$-fermion vertex (\ref{ver1}), while another
corresponds to the pole diagram with the longitudinal $a$-boson exchange
between Lorentz violating$\ a^{3}$ vertex (\ref{ver2}) and ordinary $a$%
-boson-fermion one (\ref{ver3}). Since ingoing and outgoing $a$-bosons
appear transverse ($k_{1}\cdot \epsilon ^{i}(k_{1})=0$, $k_{2}\cdot \epsilon
^{j}(k_{2})=0$) only the third term in this $a^{3}$ coupling (\ref{ver2})
contributes to the pole diagram so that one comes to a simple matrix element
$i\mathcal{M}$ for both of diagrams
\begin{equation}
i\mathcal{M}=i\frac{gn^{2}}{M}\bar{u}(p_{2})\tau ^{l}\left[ (\gamma \cdot
\mathfrak{n}^{l})+i(k\cdot \mathfrak{n}^{l})\gamma ^{\mu }k^{\nu }D_{\mu \nu
}(k)\right] u(p_{1})[\epsilon (k_{1})\cdot \epsilon (k_{2})]  \label{matr1}
\end{equation}%
where the spinors $u(p_{1,2})$ and polarization vectors $\epsilon _{\mu
}^{i}(k_{1})$ and $\epsilon _{\mu }^{j}(k_{2})$ stand for the ingoing and
outgoing fermions and $a$-bosons, respectively, while $k$ is the 4-momentum
transfer $k=p_{2}-p_{1}=k_{1}-k_{2}$. Upon the further simplifications in
the square bracket related to the explicit form of the $a$ boson propagator $%
D_{\mu \nu }(k)$ (\ref{prop1}) and matrix $\mathfrak{n}_{\mu }^{i}$ (\ref{id}%
), and using the fermion current conservation $\bar{u}(p_{2})(\hat{p}_{2}-%
\hat{p}_{1}){u(p_{1})=0}$, one is finally led to the total cancellation of
the Lorentz violating contributions to the $a$-boson-fermion scattering in
the $g/M$ approximation.

Note, however, that such a result may be in some sense expected since from
the SLIV point of view the lowest order $a$-boson-fermion scattering
discussed here is hardly distinct from the photon-fermion scattering
considered in the nonlinear QED case\cite{nambu}. Actually, the fermion
current conservation which happens to be crucial for the above cancellation
works in both of cases, whereas the couplings being peculiar to the
Yang-Mills theory have not yet touched on. In this connection the next
example seems to be more instructive.

\subsection{Vector-vector scattering}

The matrix element for this process in the lowest order $g/M$ is given by
the contact SLIV $a^{4}$ vertex (\ref{ver4}) and the pole diagrams with the
longitudinal $a$-boson exchange between the ordinary$\ a^{3}$ vertex (\ref%
{ver0}) and Lorentz violating $a^{3}$ one (\ref{ver2}), and vice versa.
There are six pole diagrams in total describing the elastic $a-a$ scattering
in the $s$- and $t$-channels, respectively, including also those with an
interchange of identical $a$-bosons. Remarkably, the contribution of each of
them is exactly canceled with one of six terms appeared in the contact
vertex (\ref{ver4}). Actually, writing down the matrix element for one of
the pole diagrams with ingoing $a$-bosons (with momenta $k_{1}$ and $k_{2}$)
interacting through the vertex (\ref{ver0}) and outgoing $a$-bosons (with
momenta $k_{3}$ and $k_{4}$) interacting through the vertex (\ref{ver2}) one
has%
\begin{eqnarray}
i\mathcal{M}_{pole}^{(1)} &=&-i\frac{gn^{2}}{M}c^{ijp}\delta
^{kl}[(k_{1}-k_{2})_{\mu }g_{\alpha \beta }+(k_{2}-k)_{\alpha }g_{\beta \mu
}+(k-k_{1})_{\beta }g_{\alpha \mu }]\cdot  \notag \\
&&\cdot D_{\mu \nu }^{pq}(k)g_{\gamma \delta }k_{\nu }(n^{q}\cdot
k)[\epsilon ^{i,\alpha }(k_{1})\epsilon ^{j,\beta }(k_{2})\epsilon
^{k,\gamma }(k_{3})\epsilon ^{l,\delta }(k_{4})]  \label{pole}
\end{eqnarray}%
where polarization vectors $\epsilon ^{i,\alpha }(k_{1})$, $\epsilon
^{j,\beta }(k_{2})$, $\epsilon ^{k,\gamma }(k_{3})$ and $\epsilon ^{l,\delta
}(k_{4})$ belong, respectively, to ingoing and outgoing $a$-bosons, while $%
k=-(k_{1}+k_{2})=k_{3}+k_{4}$ according to the momentum running in the
diagrams taken above. Again, as in the previous case of vector-fermion
scattering, due to the fact that outgoing $a$-bosons appear transverse ($%
k_{3}\cdot \epsilon ^{k}(k_{3})=0$ and $k_{4}\cdot \epsilon ^{l}(k_{4})=0$),
only the third term in the Lorentz violating $a^{3}$ coupling (\ref{ver2})
contributes to this pole diagram. Upon evident simplifications related to
the $a$-boson propagator $D_{\mu \nu }(k)$ (\ref{prop1}) and matrix $%
\mathfrak{n}_{\mu }^{i}$ (\ref{id}) one comes to the expression which is
exactly cancelled with the first term in the contact SLIV vertex (\ref{ver4}%
) when it is properly contracted with $a$-boson polarization vectors.
Likewise, other terms in this vertex provide the further one-to-one
cancellation with the remaining pole matrix elements $i\mathcal{M}%
_{pole}^{(2-6)}$. So, again, the Lorentz violating contribution to the
vector-vector scattering is absent in Goldstonic Yang-Mills theory in the
lowest $g/M$ approximation.

\subsection{Other processes}

Other tree level Lorentz violating processes, related to $a$ bosons and
fermions, appear in higher orders in the basic SLIV parameter $1/M$. They
come from the subsequent expansion of the effective Higgs mode (\ref%
{constr''}) in the Lagrangian (\ref{nab3}). Again, their amplitudes are
essentially determined by an interrelation between the longitudinal $a$%
-boson exchange diagrams and the corresponding contact $a$-boson interaction
diagrams which appear to cancel each other thus eliminating physical Lorentz
violation in theory.

Most likely, the same conclusion can be derived for SLIV\ loop contributions
as well. Actually, as in the massless QED case considered earlier \cite{ac},
the corresponding one-loop matrix elements in Goldstonic Yang-Mills theory
either vanish by themselves or amount to the differences between pairs of
the similar integrals whose integration variables are shifted relative to
each other by some constants (being in general arbitrary functions of
external four-momenta of the particles involved) that in the framework of
dimensional regularization leads to their total cancellation.

So, the Goldstonic vector field theory (\ref{nab3}) for a non-Abelian
charge-carrying matter is likely to be physically indistinguishable from a
conventional Yang-Mills theory.

\section{Conclusion}

The spontaneous Lorentz violation in 4-dimensonal flat Minkowskian
space-time was shown to generate vector Goldstone bosons both in Abelian and
non-Abelian theories with the corresponding nonlinear vector field
constraint (\ref{cons1}) or (\ref{CON}) imposed. In the Abelian case such a
massless vector boson is naturally associated with photon. In non-Abelian \
case, although the pure \ Lorentz violation still generates only one genuine
Goldstone vector boson, the accompanying vector PGBs related to a violation
of the larger accidental symmetry $SO(D,3D)$ of the SLIV constraint (\ref%
{CON}) in itself come also into play in the final arrangement of the entire
Goldstone vector field multiplet of the internal symmetry group $G$.
Remarkably, they remain strictly massless being protected by the gauge
invariance of the Yang-Mills theory involved. These theories, both Abelian
and non-Abelian, while being essentially nonlinear in the Goldstone vector
modes, are physically indistinguishable from conventional QED and Yang-Mills
theory. One could actually see that just the gauge invariance not only
provides these theories to be free from the unreasonably large Lorentz
violation stemming from the fermion and vector field bilinears (see Sections
2 and 3), but also render all the other physical SLIV effects (including
those which are suppressed by the Lorentz violation scale $M$)
non-observable (Section 4). As a result, Abelian and non-Abelian SLIV theory
appear, respectively, as standard QED and Yang-Mills theory taken in the
nonlinear gauge (to which the vector field constraints (\ref{cons1}) and (%
\ref{CON}) are virtually reduced), while the $S$-matrix remains unaltered
under such a gauge convention.

So, while at present the Goldstonic nature of gauge fields, both Abelian and
non-Abelian, seems to be highly plausible, the most fundamental question of
physical Lorentz violation in itself, that only could uniquely point toward
such a possibility, is still an open question. Note, that here we are not
dealing with direct (and quite arbitrary in essence) Lorentz non-invariant
extensions of QED or Standard Model which were intensively discussed on
their own in recent years [6-8]. Rather, the case in point is a construction
of genuine SLIV models which would generate gauge fields as the proper
vector Goldstone bosons, from one hand, and could lead to observed Lorentz
violating effects, from the other. In this connection, somewhat natural
framework for physical Lorentz violation to occur would be a model where \
the internal gauge invariance were slightly broken at very small distances
through some high-order operators stemming from the gravity-influenced area.
Such physical SLIV effects would be seen\ in terms of powers of ratio $%
M/M_{Pl}$ (where $M_{Pl}$ is the Planck mass). So, for the SLIV scale
comparable with the Planck one they would become directly observable.
Remarkably enough, if one has such internal gauge symmetry breaking in an
ordinary Lorentz invariant theory this breaking appears vanishingly small at
laboratory being properly suppressed by the Planck scale. However, the
spontaneous Lorentz violation would render it physically significant: the
higher Lorentz scale, the greater SLIV\ effects observed. If true, it would
be of particular interest to have a better understanding of the internal
gauge symmetry breaking mechanism that brings out the spontaneous Lorentz
violation at low energies. We return to this basic question elsewhere.

\section*{Acknowledgments}

We would like to thank Colin Froggatt, Rabi Mohapatra and Holger Nielsen for
useful discussions and comments. One of us (J.L.C.) is grateful for the warm
hospitality shown to him during a visit to Center for Particle and String
Theory at University of Maryland where part of this work was carried out.

\end{document}